# Interpretable deep learning illuminates multiple structures fluorescence imaging: a path toward trustworthy artificial intelligence in microscopy.

**Running title:** Interpretable Multi-structure Synchronous Fluorescence Imaging

Mingyang Chen[1,#], Luhong Jin[1,#], Xuwei Xuan[1], Defu Yang[1], Yun Cheng[2,*], Ju Zhang[1,*]

[1]School of Information Science and Technology, Hangzhou Normal University, Hangzhou, China

[2]Department of Medical Imaging, Zhejiang Hospital, Hangzhou, China. (962110508@qq.com)

[#] These authors contributed equally.

[*] to whom correspondence should be addressed. e-mails: 962110508@qq.com (Yun Cheng), juzhang@hznu.edu.cn (Ju Zhang)

**Abstract**

Live-cell imaging of multiple subcellular structures is essential for understanding subcellular dynamics. However, current techniques often require multiple rounds of staining, leading to photobleaching and reduced dye stability. Here, we present the Adaptive Explainable Multi-Structure Network (AEMS-Net), a deep-learning framework that enables simultaneous prediction of two subcellular labels from a single image acquisition. The model normalizes staining intensity and prioritizes critical image features by integrating attention mechanisms and brightness adaptation layers. Leveraging the Kolmogorov-Arnold representation theorem, our model decomposes learned features into interpretable univariate functions, enhancing the explainability of complex subcellular morphologies. We demonstrate that AEMS-Net tracks dynamic changes in mitochondrial morphology during cell migration, requiring only half the conventional staining procedures. Notably, this approach achieves over 30% improvement in imaging quality compared to traditional deep learning methods, establishing a new paradigm for long-term, interpretable live-cell imaging that advances the ability to explore subcellular dynamics.

**Introduction**

The task of multi-structural observation in live-cell microscopy imaging plays a critical role in subcellular biology. It enables researchers to investigate interactions between different types of intracellular structures, such as microtubules and mitochondria. This capability is essential for understanding disease mechanisms[1], identifying potential therapeutic targets[2], and drug screening[3]. Traditional methods enable the observation of interactions between multiple organelles in cells by combining fluorescent probes with distinct emission wavelengths and multi-color fluorescence microscopy.

However, existing multichannel fluorescence microscopy has several significant limitations. These include spectral overlap[4,5], which can cause signal crosstalk between channels, and photobleaching[6,7], which limits imaging duration. Additionally, phototoxicity[8,9] from long-term light exposure can damage live cells. Furthermore, the large volume of data generated requires complex computational analysis, and the high costs and technical demands of multichannel systems can be prohibitive. These challenges emphasize the need for continued imaging and data processing advancements. Deep learning has emerged as a promising research frontier in fluorescence microscopy imaging, offering novel approaches to overcome traditional imaging limitations. Despite its potential, the inherent black-box nature[10] of these models introduces critical challenges regarding interpretability and reliability[11], which remain key concerns in biological research applications.

Previous research has focused on specific applications of deep learning in microscopic imaging, addressing various concrete challenges in this field. For instance, Jin et al.[12] enhanced Structured Illumination Microscopy (SIM) by reducing the required raw images, thereby minimizing photobleaching effects. They also proposed a structure separation model to extract multiple distinct structures with identical fluorescent labels[13]. While Jin et al. attempted single-staining structure separation, their approach requires an additional brightness adjustment, which adds complexity. Liao et al[14]. developed a deep convolutional neural network (CNN) for direct mapping from raw data to super-resolution images, leveraging computational advantages to accelerate reconstruction and address phototoxicity issues. Many researchers have adopted U-Net[15] as their foundational framework in bioinformatics for various applications[16–22], including cell segmentation, super-resolution imaging, and microscopy enhancement, while others have explored alternative deep learning approaches such as CNN-LSTM[23], GANs[24], Transformers[25] and Diffusion[26] models.

The rapid advancement of deep learning has raised questions and concerns regarding the black-box nature of its models. In natural image processing, Zhou et al[27]. introduced a passive interpretability method through class activation mapping to elucidate model behavior. As research progressed, algorithms such as Grad-CAM[28], Grad-CAM++[29], and Layer-CAM[30] were developed and widely adopted for interpretability analysis. These advancements highlight the ongoing need for effective interpretability in deep learning models. In recent years, researchers across various fields have continued to unravel the inner workings of deep learning through interpretability studies, achieving significant results[31–38]. However, these studies have not directly addressed the concerns of cell biologists regarding the application of deep learning techniques. In other words, research on interpretable deep learning methods within the field of cell biology remains relatively scarce. Moreover, regarding the black-box model issue, previous works either lack interpretability analysis or comparative experiments for model reliability assessment.

In this work, we present a transparent deep learning framework that resolves key technical limitations in multi-structure fluorescence microscopy imaging. Our specialized deep learning approach achieves a 50% reduction in multi-staining side effects while preserving image fidelity and offering mechanistic insights into the processing pipeline. Building upon the mathematical foundations of Kolmogorov-Arnold Networks (KAN)[39], we implement its core principles within an optimized U-Net architecture. To address the technical challenges of signal intensity variations and structural heterogeneity in fluorescence imaging, we developed a specialized intensity normalization layer and an efficient attention mechanism. We demonstrate that analyzing successful and failed cases in interpretability studies strengthens the reliability assessment of deep learning models. The quantitative evaluation shows that our method consistently surpasses conventional U-Net implementations across multiple performance metrics. Additionally, we modified the Layer-CAM algorithm to enable detailed interpretability analysis of our model. This advancement gives researchers a precise mechanistic understanding of multi-structure reconstruction processes and establishes a robust foundation for explainable deep learning in microscopy.

# Results

**Enhanced Performance through Architectural Innovation**

Previous contents have examined the drawbacks of multiple staining rounds. Here, we present subcellular microscopy images comparing double-staining versus single-staining approaches (Fig. 1a). Our goal centered on reconstructing mitochondria and microtubules through a single staining procedure, which would not only streamline laboratory protocols but also preserve subcellular viability - establishing a more robust foundation for live-cell imaging. To achieve this, we integrated the Kolmogorov-Arnold representation theorem into U-Net architecture, resulting in an innovative deep learning framework: AEMS-Net (Fig. 1b). This neural network incorporates KAN convolution[40], attention mechanisms, and adaptive brightness layer to reconstruct multi-subcellular structures. Within AEMS-Net, we initially overlay mitochondrial and microtubular structures from double-staining images while feeding the original image as paired input to the network (Fig. 1c). This approach ensures that AEMS-Net effectively captures both structural features. The network then processes these inputs through multiple KAN convolution layers (Fig. 1d, Supplementary Fig. S3) with downsampling operations (Supplementary Fig. S1), preserving multi-scale features that undergo enhancement through attention modules. During subsequent feature fusion, original features enter through residual connections[41], maintaining stable gradient flow throughout model training. To address brightness variations arising from staining overlay, we introduced an adaptive brightness pooling layer (Fig. 1e) after the final up sampling operation, enabling AEMS-Net to minimize brightness-related interference in separation and reconstruction tasks.

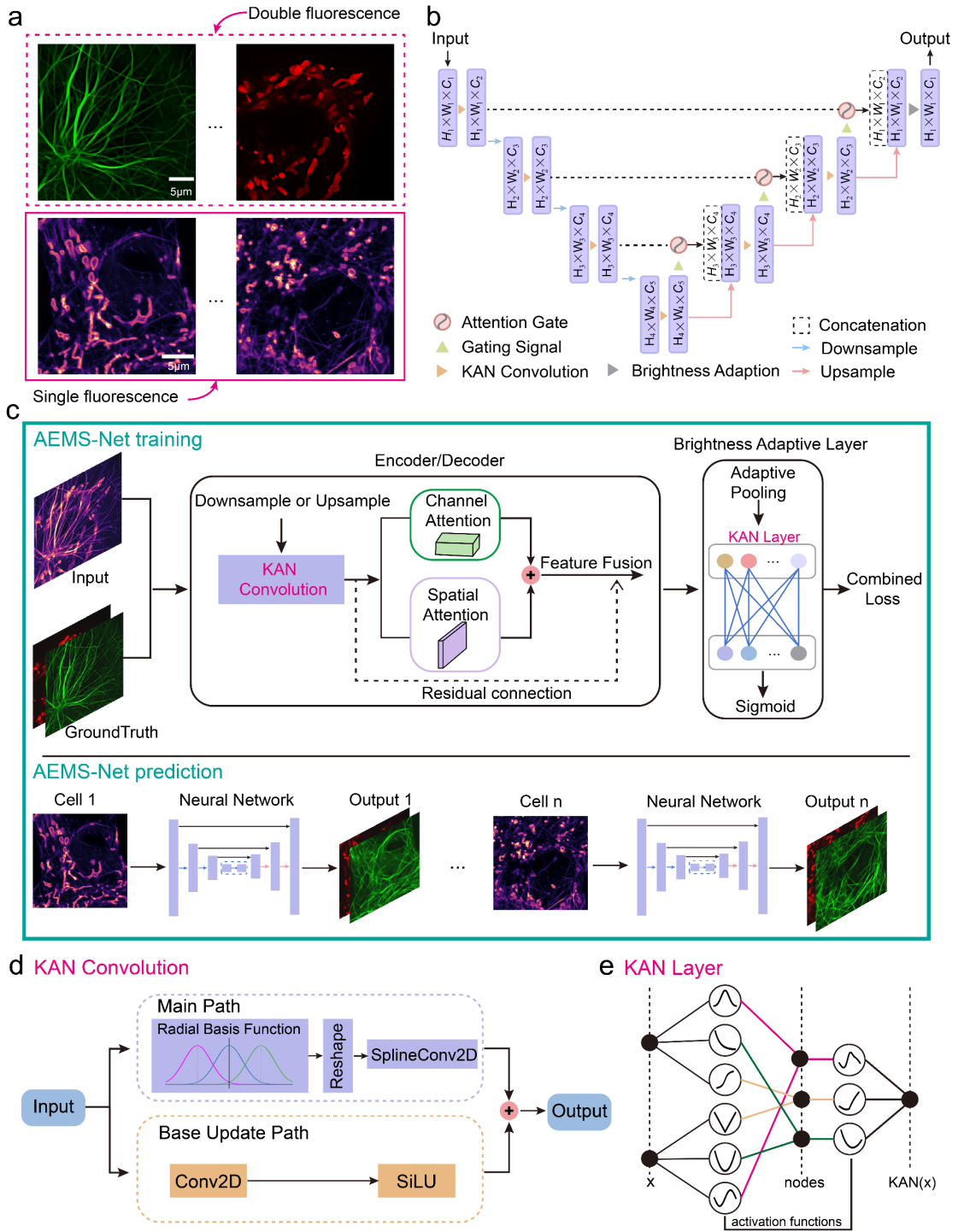

**Fig. 1 AEMS-Net Architecture Overview. a Dataset Overview.** The dashed boxes indicate subcellular structures requiring dual staining, while solid boxes represent those needing single staining. Traditional approaches necessitate double staining for observing two structures, whereas our method achieves visualization through a single staining procedure. **b AEMS-Net Architecture.** Input images undergo downsampling operations utilizing KAN convolution, which reduces resolution by half at each step while performing attention activation. During up sampling, feature maps establish residual connections. The final layer implements brightness adaptation, generating separated reconstruction images. **c AEMS-Net Workflow.** During training process, the network receives both the overlaid images of mitochondria and microtubules from single staining and their corresponding ground truth as image pairs. Throughout down sampling and up sampling, the process incorporates KAN convolution followed by attention fusion modules. Each step concludes with brightness adaptation before outputting results. The network compares outputs against ground truth using combined loss for backpropagation. During prediction process, single-stained images enter AEMS-Net to generate separate mitochondrial and microtubular images. **d KAN Convolution Process.** The input bifurcates into two branches. The first branch applies Radial Basis Function transformation, reshaping data to accommodate Spline convolution dimensions, initialized with piecewise normal distribution. The second branch employs standard 2D convolution activated through SiLU function - a smooth, non-monotonic activation mechanism. The process culminates in the concatenation of both branch outputs. **e KAN Layer Illustration.** Unlike Multilayer Perceptron (MLP) which utilizes fixed activation functions and learns weights, KAN implements addition at nodes while learning activation functions at edges. All activation functions maintain non-linear learning capabilities, culminating in their summation.

We partitioned the superimposed images into training, validation, and test sets (Fig 2a, 2b, 2c), with rigorous measures to prevent cell information leakage between these divisions (Supplementary Table 1, Table 2). Through extensive evaluations comparing AEMS-Net against a structurally identical U-Net, we conducted comprehensive analyses using Peak Signal-to-Noise Ratio (PSNR), Normalized Root Mean Square Error (NRMSE), Structural Similarity Index (SSIM) metrics (Fig2d, 2e), reconstructed line profile analysis (Fig2f), and qualitative reconstruction assessment. For microtubules, U-Net incorrectly classified mitochondria (Fig. 2d region 1, 2, yellow arrows). Compared to ground truth (Fig2b region 1, 2, yellow arrows), AEMS-Net achieved markedly superior alignment with actual values (Fig2e region 1, 2, yellow arrows), demonstrating exceptional microtubule extraction capabilities. This misclassification naturally led to incomplete mitochondria reconstruction. The experimental results provided compelling evidence for this phenomenon. U-Net generated mitochondrial images exhibited notable omissions (yellow arrows, Fig. 2d, regions 3 and 4), as these structures were erroneously segregated into the microtubule channel. In contrast, AEMS-Net accurately captured these challenging regions (Fig2e region 3, 4, yellow arrows), displaying remarkable discrimination ability. Further analysis through line profile plots revealed the reconstruction performance of both networks (Fig2f). AEMS-Net demonstrated superior alignment with original data (orange dashed line, blue solid line). Comprehensive testing across the entire test dataset (Fig2g, Supplementary Table 3) reinforced AEMS-Net as the superior method, exhibiting exceptional performance and reliability.

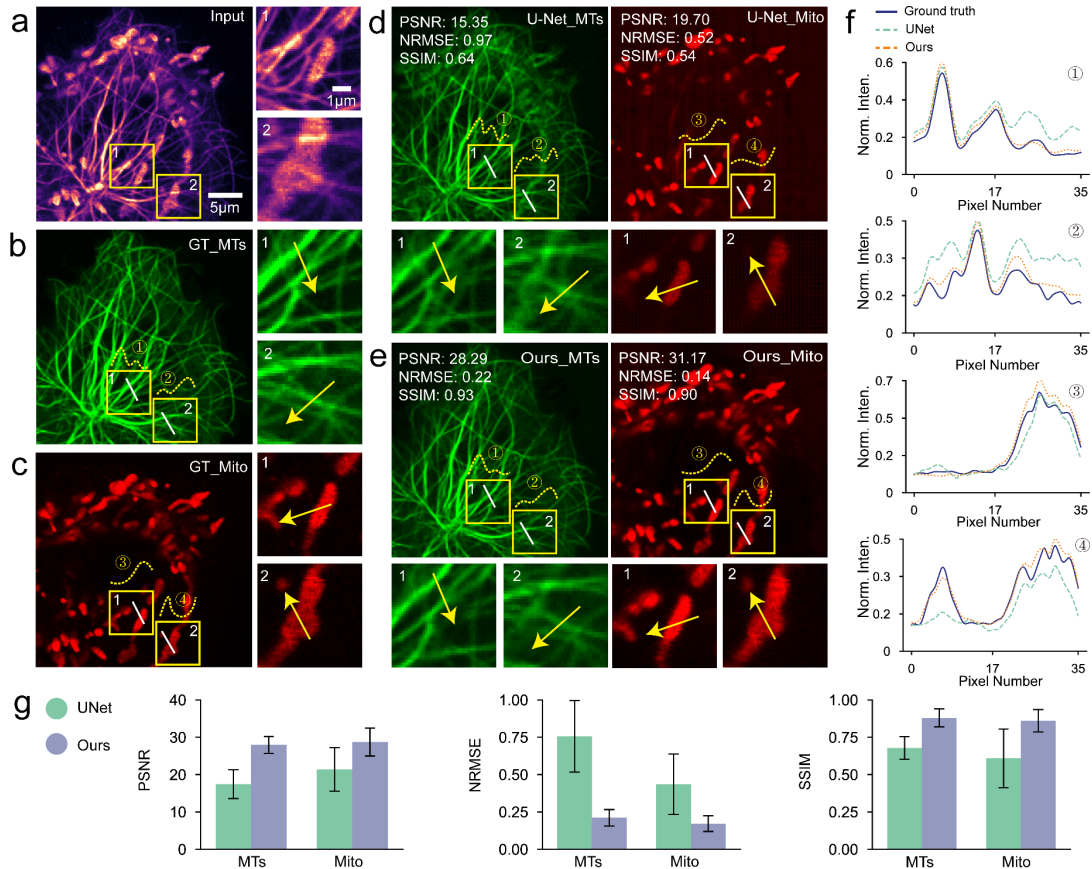

**Fig. 2 Performance evaluation of AEMS-Net on test datasets. a** Merged visualization of mitochondria and microtubules in test images, with two regions of interest (yellow boxes) highlighted and magnified. Yellow arrows indicate key features. **b-c** Ground truth images of microtubules and mitochondria, with line profiles (designated as lines 1-4) extracted from selected regions to obtain raw intensity distributions of microtubules. **b microtubules (MTs) and c mitochondria (Mito)**. **d-e** Reconstruction outcomes from U-Net and AEMS-Net. In regions 1 and 2, U-Net incorrectly categorizes mitochondrial structures as microtubules (yellow arrows), resulting in missing mitochondrial reconstructions in regions 3 and 4 (yellow arrows). Our method demonstrates superior fidelity to ground truth. **f** Comparative analysis of normalized intensity profiles across four regions. U-Net results (green dashed line) exhibit substantial deviations from raw data with classification errors in specific ranges, whereas our method (orange dashed line) demonstrates enhanced alignment with original distributions. **g** Structure-specific quantitative assessment through PSNR, NRMSE, and SSIM metrics, with error bars representing standard error of mean (SEM). Numerical evidence reveals superior performance of our method in both metric scores and stability. All raw data utilized are provided as Source Data files.

We conducted qualitative evaluations of AEMS-Net beyond standard test sets. The network processed dual-structure fluorescence subcellular images from single staining, followed by brightness and contrast normalization. Our analysis focused on two key aspects: extraction efficiency of individual structures (Fig. 3) and capture of dynamic interaction processes across time series (Fig. 4, Supplementary Fig.S5). For individual structures, our goal was to achieve optimal separation of mitochondria and microtubules, enabling researchers to conduct detailed downstream analyses. U-Net began losing critical microtubule details at multiple time points (5s, 75s) (Fig. 3b region 1, yellow arrows). In contrast, AEMS-Net preserved these subtle features (Fig. 3c region 1, yellow arrows), even when the original subcellular fluorescence images showed weak microtubule intensity (Fig. 3a region 1, yellow arrows)—the same principle held for mitochondrial extraction. At later time points (180s, 245s), U-Net incorrectly interpreted mitochondrial structures as fragmented (Fig. 3d region 2, yellow arrows), despite their continuous nature in reality (Fig. 3a, region 2, yellow arrows). AEMS-Net maintained fidelity to the actual biological phenomena (Fig. 3e, region 2, yellow arrows) - a crucial requirement for rigorous biological research.

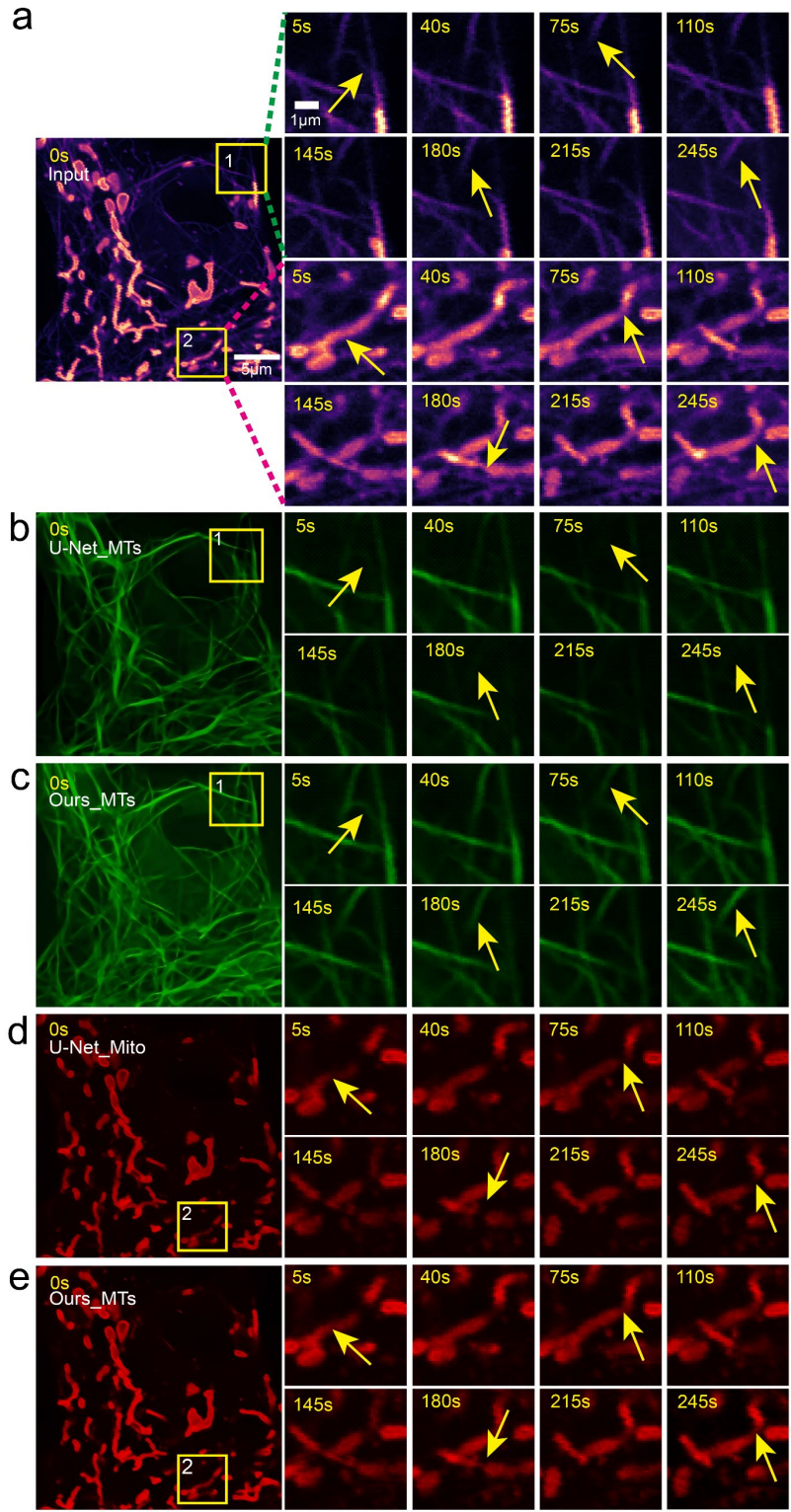

**Fig. 3 Comparative analysis of structural consistency between AEMS-Net and U-Net across temporal sequences. a** Representative time points from the dataset highlight two regions of interest (yellow boxes), with arrows indicating crucial performance distinctions between AEMS-Net and U-Net. **b-c** Microtubule reconstruction sequences demonstrate the comparative performance of U-Net versus AEMS-Net. At 5s, 75s, 180s, and 245s, AEMS-Net exhibits superior preservation of microtubule architecture compared to U-Net (yellow arrows),

maintaining structural fidelity across temporal points. **d-e** Mitochondrial reconstruction sequences reveal the differential capabilities of both networks. At 75s and 180s, AEMS-Net accurately preserves mitochondrial morphology, whereas U-Net erroneously indicates mitochondrial fragmentation where none exists (yellow arrows in subfigure **a** at 75s and 180s). The complete temporal sequence appears in Supplementary Video 1.

Beyond examining isolated structural changes, the intricate interplay between subcellular components reveals fundamental biological phenomena. To capture these dynamic interactions, we merged AEMS-Net output channels while normalizing the fluorescence intensities of microtubules and mitochondria. Mitochondria navigate along microtubule networks, orchestrating essential fission and fusion dynamics. At 145s and 215s, our imaging revealed mitochondria crawling along microtubule networks (yellow arrows, Fig. 4a). While U-Net incorrectly interpreted these events as mitochondrial fragmentation (yellow arrows, Fig. 4b), AEMS-Net accurately reconstructed the authentic biological behavior (yellow arrows, Fig. 4c). To facilitate detailed analysis of these intricate dynamics, we provide time-lapse recordings from two distinct living cells (Supplementary Videos 1 and 2).

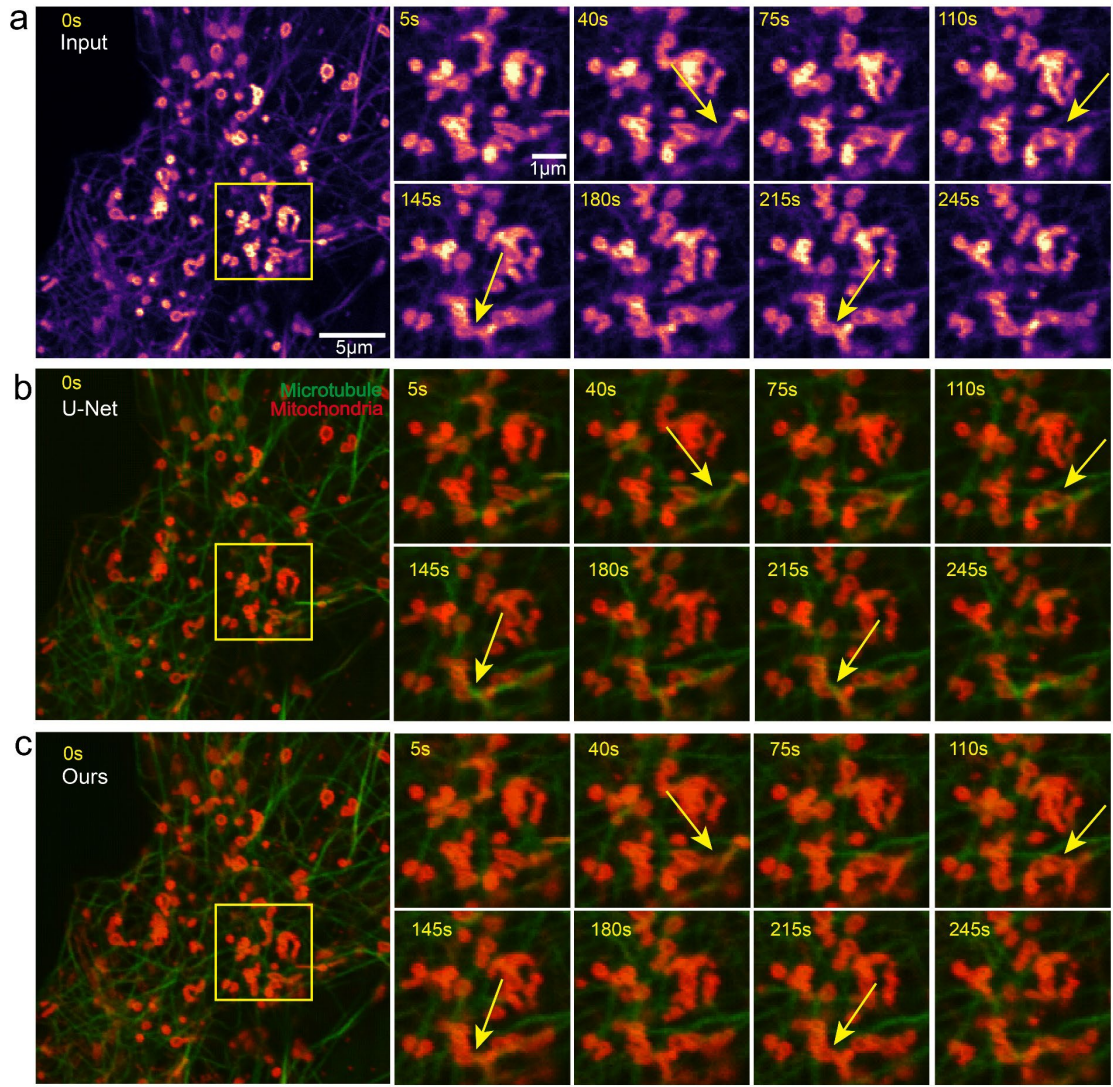

**Fig. 4 Live-cell application revealed the high efficiency of our AEMS-Net in tracking the dynamic interactions. a** Time-lapse live-cell image sequences with a selected region (yellow box) for analyzing dynamic interactions between two subcellular structures. Arrows indicate the biological events of interest. **b-c** Output comparisons between U-Net and AEMS-Net across the time series. At 40s, 110s, 145s, and 215s, U-Net failed to capture mitochondrial movement along microtubules and incorrectly interpreted mitochondrial fission events that did not occur (yellow arrows in subfigure a). AEMS-Net demonstrated enhanced accuracy in tracking subcellular structure interactions, with arrow-indicated events closely matching those in panel a. The complete time-lapse sequence can be found in Supplementary Video 2. To ensure fair visual comparison between the two models and eliminate potential contrast bias from different output intensities, we normalized the contrast of both model outputs according to the original input intensity.

We conducted a series of ablation studies to assess the contribution of individual components in AEMS-Net, focusing on the Attention module and Brightness Adaptation Layer (BAL). We quantified the performance metrics under different ablation conditions to demonstrate the impact of each component on model performance (Table 2). The Attention and BAL modules combined yielded the most significant improvements in PSNR, NRMSE, and SSIM scores. For instance, incorporating both modules increased the PSNR for mitochondria reconstruction from 23.11 ± 3.65 to 28.03 ± 3.50 and for microtubules from 21.2 ± 1.72 to 28.25 ± 2.64. Similarly, SSIM values improved from 0.64 ± 0.15 to 0.83 ± 0.09 for mitochondria and from 0.68 ± 0.08 to 0.87 ± 0.06 for microtubules. The experimental findings demonstrate the complementary interaction between the Attention mechanism and Brightness Adaptive Layer (BAL), which collectively enhance the performance of AEMS-Net. This synergistic integration proves particularly effective in maintaining intricate structural details and elevating image quality across diverse subcellular constituents. The observed improvements manifest through enhanced preservation of morphological features and superior delineation of subcellular components.

**Table 2 Ablation Study of Attention Modules in Image Restoration**

| Our method | | PSNR | | NRMSE | | SSIM | |
|---|---|---|---|---|---|---|---|
| Attention | BAL | Mito | MTs | Mito | MTs | Mito | MTs |
| 0 | 0 | 23.11±3.65 | 21.2±1.72 | 0.33±0.1 | 0.43±0.1 | 0.64±0.15 | 0.68±0.08 |
| 0 | 1 | 21.69±4.73 | 20.93±2.06 | 0.4±0.15 | 0.45±0.11 | 0.69±0.15 | 0.71±0.09 |
| 1 | 0 | 27.1±2.7 | 28.01±2.44 | 0.2±0.05 | 0.2±0.06 | 0.81±0.09 | 0.84±0.07 |
| 1 | 1 | 28.03±3.5 | 28.25±2.64 | 0.18±0.05 | 0.2±0.06 | 0.83±0.09 | 0.87±0.06 |

\* Mito: Mitochondria, MTs: Microtubule

To rigorously evaluate model robustness during quantitative analysis, we deliberately scaled specific structural images within reasonable ranges when preparing the training dataset. For example, we applied proportional magnification to microtubule images - a common data augmentation technique in deep learning that enhances model reliability. This modification aligns well with real-world applications, where microtubules and mitochondria naturally exhibit dimensional variations. Our comprehensive experimental data demonstrate that AEMS-Net exhibits minimal sensitivity to subcellular structure dimensions, maintaining focus on morphological characteristics for segmentation and reconstruction. In contrast, U-Net performance deteriorated with dimensional changes (Supplementary Fig.S4, Supplementary Table 4, Table 5).

In sum, our method achieves superior performance in live-cell applications. The improvements in accuracy, generalization, and efficiency allow for more detailed and insightful analyses of dynamic interactions between subcellular structures.

**Interpretability and Transparency in AEMS-Net**

Explainable Artificial Intelligence (XAI)[42] emerged as a pivotal research direction, particularly in biomedical applications where researchers require both exceptional performance and reliability from deep learning models. To enhance the interpretability of AEMS-Net, we visualized the model decision steps through gradient flow analysis (Fig. 5a). We refined the Layer-CAM algorithm (Supplementary Fig.S2) to achieve optimal compatibility with AEMS-Net. The process involved preserving feature maps during encoding and decoding while computing gradients through backpropagation. To thoroughly validate AEMS-Net interpretability, we developed and implemented forward and reverse activation algorithms. When examining microtubule separation mechanisms, we analyzed critical and non-critical features (Supplementary Fig. S6-S11). The complementary nature between positive activations and least-considered features enhanced decision transparency, demystifying the black-box nature of the model and bolstering user confidence. Our comparative analysis between U-Net and AEMS-Net examined each down sampling and up sampling module through activation heatmap visualization (Fig. 5b). The interpretability analysis of the Up3 module revealed that AEMS-Net precisely captured the elongated morphological characteristics of microtubules (Fig. 5b, Ours_MTs, Up2, Up3). For mitochondria, it accurately identified their spherical and punctate morphological features (Fig. 5b, Ours_Mito, Up2, Up3). In contrast, U-Net exhibited broad, unfocused attention patterns and failed to distinguish effectively between mitochondria and microtubules (Fig. 5g, U-Net_MTs, U-Net_Mito, Up2, Up3), explaining the separation errors observed (Fig. 2d region 1-4). To our knowledge, Previous studies[31,43–46] have yet to delve into such a detailed analysis of profound learning model limitations in this context. Our interpretability analysis not only elucidates why AEMS-Net excels but also reveals why comparative models fall short, offering practical value for developing reliable artificial intelligence in bioinformatics.

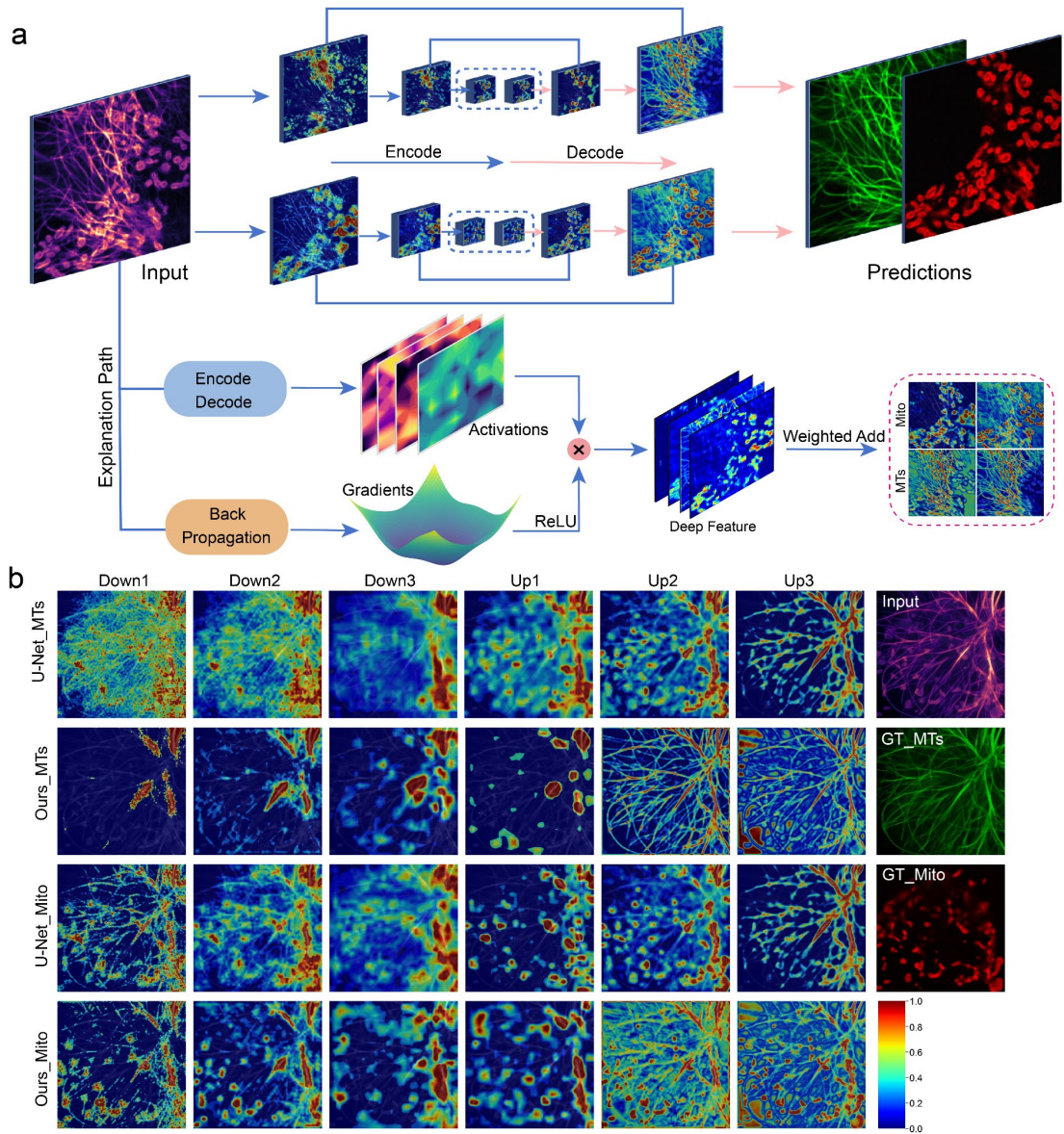

**Fig. 5 Interpretability Analysis of AEMS-Net. a** Heatmap visualization of pixel importance and interpretable algorithm workflow. During training, AEMS-Net analyzes decisions at each step based on intrinsic features of different subcellular structures. This process becomes visible through heatmaps, enabling researchers to understand the deep learning model reasoning. The forward propagation preserves activation maps, while backpropagation calculates gradients. The computed gradients underwent ReLU activation, which truncates negative values to zero. To address the resultant information loss, we incorporated inverse gradient analysis (Supplementary Fig. S2), revealing regions that received minimal attention from the deep learning model. This bidirectional interpretability approach enabled comprehensive examination of both highlighted and overlooked features. **b** Interpretability analysis comparing U-Net and AEMS-Net across different modules. During downsampling, U-Net and AEMS-Net demonstrate distinct strategies. U-Net maintains broad attention regions when processing both mitochondria and microtubules (U-Net_MTs Down1, U-Net_Mito Down1). In contrast, AEMS-Net adapts attention based on structural characteristics - focusing on stem-like features for microtubules (Ours_MTs Down1) while shifting to broader, punctate patterns for mitochondria (Ours_Mito Down1). Furthermore, during the final two up sampling steps, when structural reconstruction should near completion, U-Net exhibits unclear decision-making for microtubule separation, shown by punctate heatmaps (U-Net_MTs Up2). AEMS-Net, however, presents clear justification for microtubule reconstruction through heatmaps that align with their thin, tubular characteristics (Ours_MTs Up2). U-Net interpretability maps reveal minimal distinction between mitochondria and microtubules (U-Net_MTs Up3, U-Net_Mito Up3), despite their evident structural differences. AEMS-Net reveals robust mechanistic principles for organelle separation and reconstruction. The interpretability analysis demonstrates distinct yet complementary spatial patterns between mitochondrial and microtubular structures (Ours_MTs Up3, Ours_Mito Up3), reflecting computational processes that align with expert biological reasoning.

The experimental results demonstrate that AEMS-Net exhibits exceptional efficiency in extracting diverse subcellular structures through well-established mechanistic principles. This deep learning architecture demonstrates robust capabilities in learning distinguishing characteristics of distinct subcellular components, thereby ensuring reliable performance in practical biological applications. Interpretability analyses conducted on single-stained subcellular specimens yielded findings that exhibited consistent alignment with the experimental validation outcomes (Supplementary Fig. S11).

## Discussion

Multiple fluorescence staining introduces adverse effects for live cell imaging, including spectral crosstalk and phototoxicity. While deep learning approaches have alleviated these challenges, they remain fundamentally limited by poor interpretability. The advancement of subcellular imaging proceeds along two parallel trajectories: developing novel live-cell and subcellular fluorescence techniques, and enhancing established methodologies. Within this broader context, our research focuses on two challenges: circumventing the limitations inherent in multiple staining protocols while establishing transparent deep learning frameworks that foster confidence in AI-driven biomedical applications.

In this work, we introduce AEMS-Net for subcellular structure separation and reconstruction. Compared to the widely adopted U-Net architecture, AEMS-Net demonstrates three distinct advantages: First, it exhibits enhanced learning and representation capabilities, capturing subcellular structural features with higher reconstruction quality scores despite limited training data. Second, AEMS-Net leverages Layer-CAM to illuminate factors influencing separation and reconstruction, quantifying pixel-level significance through heatmap visualization - an interpretability analysis applicable to any encoder-decoder network architecture. Third, AEMS-Net enables expandable training for diverse subcellular structures. For instance, the traditional analysis of five subcellular structures demands five separate fluorescence procedures, introducing complexity and risking cell death. AEMS-Net eliminates four staining steps and filters changes through its extensible learning capabilities.

Beyond these advantages, we validated AEMS-Net performance across temporal sequences (Supplementary videos 1 and 2), demonstrating real-time inference capabilities. This breakthrough enables AEMS-Net integration into event-driven intelligent microscopy systems[47,48], allowing neural network-based structured illumination microscopy control triggered by biological phenomena.

The current version of AEMS-Net allows only passive, post-hoc interpretability analysis, which hinders real-time error correction when deep learning models show bias. Comprehensive investigations into inherently interpretable artificial intelligence systems with active learning capabilities remain nascent. The predominant methodological framework integrates Bayesian probabilistic modeling[49] with advanced deep learning architectures, seeking to develop transparent AI systems that actively refine their interpretability through continuous learning processes. This approach represents a paradigm shift from post-hoc explanations toward architecturally embedded interpretability. Although passive interpretability methods do not represent an optimal solution in ideal circumstances, these approaches retain significant value during the evolution toward active interpretability. Researchers can substantially reduce trial-and-error costs and avoid directionless experimentation through such interpretative analyses. Notably, KAN complements AEMS-Net by addressing passive interpretability limitations. KAN demonstrates the ability to derive task-specific expressions through optimization operations like pruning, reducing model complexity[39,50]. Naturally, shallower model depth translates to enhanced interpretability and transparency. While experiments validate the effectiveness of KAN within AEMS-Net for the current task, the potential applications of active interpretability in image processing remain largely unexplored. This open territory, particularly regarding PDE-based approaches, defines the direction for future research.

# Methods

## Cell Culture and Preparation

The COS-7 cells were cultured in high glucose Dulbecco's modified Eagle's medium (DMEM) (Gibco, #11965092), supplemented with 10% fetal bovine serum (Sigma-Aldrich, #F8313) and 1% penicillin–streptomycin (Beyotime, #C0222) at 37°C in a humidified 5% CO2 incubator. Cells were planted into a 35-mm glass bottom dish (Cellvis, #D35-20-1-N) for fluorescence imaging experiments.

For the training, testing and validation steps, COS-7 cells were transfected with EMTB-3×eGFP and stained with MitoTracker Orange (Thermo Fisher, M7510) to label the microtubules (MTs) and mitochondria (Mito) respectively. To demonstrate the application potential, we co-transfected COS-7 cells with EMTB-3×eGFP and Tom20-mEmerald.

The images were acquired by Olympus FV3000 fluorescence microscope equipped with 488 nm and 561 nm laser lines. The objective lens is 100X (1.45 NA). The confocal imaging speed for each image is around 5s.

## Data Preprocessing

The experimental protocol involved the independent preparation and staining of mitochondrial and microtubule specimens for microscopic visualization. The corresponding subcellular architectures were meticulously integrated while maintaining rigorous segregation among training, validation, and testing cohorts. A systematic random-sampling approach was implemented to augment the dataset dimensionality and enhance the deep learning model efficacy. This methodology encompassed the extraction of $256 \times 256$-pixel regions from each subcellular ensemble, followed by comprehensive intra-group image superimposition analyses. The systematic approach generated an extensive image repository encompassing 2,142 training specimens, 117 validation specimens, and 720 test specimens (Supplementary Table 2). Although inherent variations in fluorescence intensity were observed between mitochondrial and microtubule channels, the deliberate omission of image preprocessing aimed to minimize operational complexity and resource requirements for the research community. Network performance optimization incorporated intensity normalization through maximum intensity scaling, mapping pixel values to the interval [0,1]. Subsequently, the images underwent transformation into $256 \times 256 \times c$ PyTorch tensors, wherein c denotes the channel dimensionality, facilitating neural network training procedures.

**Loss Function and Training Details**

Segmentation of complex subcellular structures requires precise loss function design that captures the intricate spatial relationships between subcellular components. We developed a unified loss function that simultaneously addresses the segmentation of mitochondria and microtubules, integrating a joint loss approach with L1 regularization to enhance structural representation and generalization. The proposed loss function is defined as:

$$L_{total} = L_{mito} + L_{micro} + \lambda \times L1_{reg} \qquad (1)$$

Where $L_{\text{mitochondria}}$ and $L_{\text{microtubules}}$ represent the respective structure-specific segmentation losses, and $\lambda \times L1_{reg}$ denotes the L1 regularization term that promotes sparsity in the model parameters.

Several equations jointly regulate the loss functions for mitochondria and microtubules, as represented by the combined loss function shown below:

$$L_{mito/micro} = w_{mse} \times L_{mse} + w_{grad} \times L_{grad} + w_c \times L_c + w_{focal} \times L_{focal} \qquad (2)$$

$$L_{mse} = \frac{1}{n}\sum_{i=1}^{n}(y_i - \hat{y}_i)^2 \qquad (3)$$

$$L_{grad} = \frac{1}{n}\sum_{i=1}^{n}(\nabla y_i - \nabla \hat{y}_i) \qquad (4)$$

$$L_{contrastive} = -log\frac{exp(sim(y_i, \hat{y}_i)/T)}{\sum_{j=1}^{n}exp(sim(y_i, \hat{y}_j)/T))} \qquad (5)$$

$$L_{focal} = -\alpha(1 - p_t)^{\gamma} \times log(p_t) \qquad (6)$$

The proposed loss function is a weighted combination that integrates four distinct sub-loss components: mean squared error loss ($L_{mse}$), gradient loss ($L_{grad}$), contrastive loss ($L_c$), and focal loss ($L_{focal}$). The weight coefficients for each component are denoted as, $w_{mse}, w_{grad}, w_c$, and $w_{focal}$ respectively.

For $L_{mse}$ (Mean Squared Error Loss), n denotes the number of samples, where $y_i$ represents the true value of the i-th sample and $\hat{y}_i$ indicates the predicted value of the i-th sample.

For $L_{grad}$ (Gradient Loss), $\nabla y_i$ denotes the gradient of the true values and $\nabla \hat{y}_i$ represents the gradient of the predicted values.

For $L_{contrastive}$ (Contrastive Loss), $sim(y_i, \hat{y}_i)$ indicates the similarity measure between the true values and the predicted values. $T$ is the temperature parameter used to adjust the smoothness of the distribution, and $exp()$ denotes the exponential function.

For $L_{focal}$ (Focal Loss), $\alpha$ is the adjustment parameter, $p_t$ represents the predicted probability value, and $\gamma$ is the focusing parameter used to adjust the weight of hard-to-classify samples.

The AEMS-Net processes input images with dimensions of 256×256 pixels. The initial preprocessing stage involves image transformation into PyTorch tensors, followed by convolution operations that expand the channel dimension to 64. Through sequential downsampling operations, the spatial dimensions undergo progressive reduction by factors of two while the channel depth doubles iteratively, ultimately generating a compact bottleneck representation of 16×16×1024. The subsequent upsampling phase reconstructs these encoded features to dimensions of 256×256×64. The final reconstruction phase incorporates a brightness adaptation layer for output refinement (Supplementary Fig.S3). The implementation of KAN convolution introduces architectural modifications wherein the ImprovedFastKAN double convolution transforms the Radial Basis Function (RBF) to generate high-dimensional feature representations. This process employs grid-based centroid modeling coupled with spline convolution for sample point fitting, thereby enhancing the nonlinear representation capabilities of the network. The complete implementation protocol and source code have been made publicly accessible through GitHub repository.

**Evaluation Metrics**

We use PSNR, NRMSE, and SSIM as evaluation metrics for the test set, and each metric can be calculated using the following formulas:

$$PSNR = 20 \times log_{10}(\frac{MAX_I}{\sqrt{\sum_{i=1}^{W}\sum_{j=1}^{H}((U(i,j)-V(i,j))^2/(W \times H)}}) \tag{7}$$

$$NRMSE = \frac{\sqrt{\sum_{i=1}^{W}\sum_{j=1}^{H}(U(i,j)-V(i,j))^2}}{\sqrt{\sum_{i=1}^{W}\sum_{j=1}^{H}(U(i,j))^2}} \tag{8}$$

$$SSIM(x,y) = \frac{(2\mu_x\mu_y + C_1)(2\sigma_{xy} + C_2)}{(\mu_x^2\mu_y^2 + C_1)(\sigma_x^2\sigma_y^2 + C_2)} \tag{9}$$

$MAX_I$ represents the maximum pixel value of the image. $U(i,j)$ denotes the pixel value of the original image at the position $(i,j)$, while $V(i,j)$ indicates the pixel value of the predicted image at the same position. W and H refer to the width and height of the image, respectively, and i and j are the index values for the pixel coordinates. Furthermore, $\mu_x$ and $\mu_y$ represent the mean values of images x and y, respectively, and $\sigma_x$ and $\sigma_y$ denote the standard deviations of images x and y. The term $\sigma_{xy}$ signifies the local covariance between images x and y. $C_1$ and $C_2$ are constant terms added to avoid division by zero. Lastly, x and y refer to the original image and the predicted image, respectively.
We used Python code to calculate these three metrics, utilizing the functions from the skimage.metrics library.(https://pypi.org/project/scikit-image)

**Generalization and Interpretability Studies**

In the present investigation, we conducted a rigorous evaluation of two deep learning architectures, AEMS-Net and U-Net, utilizing the Application dataset, which encompasses a comprehensive collection of single-stained microscopic images depicting multiple subcellular structures (Supplementary Table 1). This dataset captures complex biological dynamics, including mitochondrial motility, fusion events, and fission processes along microtubule networks across temporal dimensions.

The trained models were challenged with unmodified Application dataset inputs to evaluate their generalization capabilities on novel, previously unexamined data. Our systematic analysis focused on specific temporal intervals, examining the comparative efficacy of AEMS-Net and U-Net in two critical aspects: the extraction of individual structural elements and the reconstruction of dynamic multi-structural interactions. The experimental results demonstrated that AEMS-Net exhibited superior performance across both evaluation metrics (Fig. 3, 4, Supplementary Fig.S3, S4).

Recognizing the importance of model generalization, we employed the interpretability module of AEMS-Net to conduct a detailed analysis of the input images (Fig. 5, Supplementary Fig.S2). This process began with a forward pass to save the feature maps, followed by a backward pass to compute gradients, yielding both positive and negative activation values. The subsequent feature processing module allowed us to save the interpretable activation maps from all encoder and decoder layers (Supplementary Fig. S6-9).

Upon analyzing the interpretable activation maps of AEMS-Net and U-Net (Supplementary Fig.S11), it became evident that our method aligns more closely with human intuitive reasoning. AEMS-Net effectively delineated the slender characteristics of microtubules and the spherical nature of mitochondria. At the same time, U-Net struggled to provide such clarity, with minimal distinction between the positive and negative activation maps. This comprehensive evaluation underscores the robustness and interpretability of our method in the context of subcellular imaging analysis.

## Data and Code Availability

All relevant data are included within the article and its Source Data. Due to size limitations, the training datasets can be obtained from the corresponding author upon request. The original data for Fig. 2, Table 1, 2, and Supplementary Table 3,4, are provided as a Source Data File.

The source code and pre-trained models used in this study are available at the following GitHub link:

**Supplementary Table 1. Cell information**

| |
|---|
| Mitochondria: |
| Cell_1:   Sample_000.tif   -   Sample_008.tif |
| Cell_2:   Sample_009.tif   -   Sample_018.tif |
| Cell_3:   Sample_019.tif   -   Sample_033.tif |
| Cell_4:   Sample_034.tif   -   Sample_050.tif |
| Cell_5:   Sample_051.tif   -   Sample_058.tif |
| Cell_6:   Sample_059.tif   -   Sample_064.tif |
| Cell_7:   Sample_065.tif   -   Sample_071.tif |
| Cell_8:   Sample_072.tif   -   Sample_075.tif |
| Cell_9:   Sample_076.tif   -   Sample_081.tif |
| Cell_10:  Sample_082.tif   -   Sample_084.tif |
| |
| Microtubule: |
| Cell_1:   Sample_000.tif   -   Sample_007.tif |
| Cell_2:   Sample_008.tif   -   Sample_025.tif |
| Cell_3:   Sample_026.tif   -   Sample_045.tif |
| Cell_4:   Sample_046.tif   -   Sample_065.tif |
| Cell_5:   Sample_066.tif   -   Sample_081.tif |
| Cell_6:   Sample_082.tif |
| Cell_7:   Sample_083.tif |
| Cell_8:   Sample_084.tif |
| |
| Application |
| Cell_1:   Sample_00.tif   -   Sample_049.tif |
| Cell_2:   Sample_00.tif   -   Sample_019.tif |
| Cell_3:   Sample_00.tif   -   Sample_049.tif |
| Cell_4:   Sample_00.tif   -   Sample_049.tif |
| Cell_5:   Sample_00.tif   -   Sample_049.tif |

## Supplementary Table 2. Dataset Statistics Across Structures

|  |  | Mitochondria | Microtubule | total number |
|---|---|---|---|---|
| training | | Cell_3: Sample_019.tif - Sample_033.tif | Cell_3: Sample_026.tif - Sample_045.tif | 2142 |
| | | Cell_4: Sample_034.tif - Sample_050.tif | Cell_4: Sample_046.tif - Sample_065.tif | |
| | | Cell_6: Sample_059.tif - Sample_064.tif | Cell_6: Sample_082.tif | |
| | | Cell_7: Sample_065.tif - Sample_071.tif | Cell_7: Sample_083.tif | |
| | | Cell_9: Sample_082.tif - Sample_084.tif | / | |
| validation | | Cell_1: Sample_000.tif - Sample_008.tif | Cell_1: Sample_000.tif - Sample_007.tif | 117 |
| | | Cell_8: Sample_072.tif - Sample_075.tif | Cell_8: Sample_084.tif | |
| testing | | Cell_2: Sample_009.tif - Sample_018.tif | Cell_2: Sample_008.tif - Sample_025.tif | 720 |
| | | Cell_5: Sample_051.tif - Sample_058.tif | Cell_5: Sample_066.tif - Sample_081.tif | |
| | | / | Cell_10: Sample_082.tif - Sample_084.tif | |

**Supplementary Table 3. Quantitative estimation of U-Net and our method after scaling**

| method | metric | psnr_mito | psnr_micro | nrmse_mito | nrmse_micro | ssim_mito | ssim_micro |
|---|---|---|---|---|---|---|---|
| U-Net | avg±std | 21.38±5.86 | 17.45±3.87 | 0.44±0.2 | 0.76±0.3 | 0.61±0.2 | 0.68±0.08 |
| | max | 32.63 | 27.59 | 0.77 | 1.37 | 0.89 | 0.84 |
| | min | 12.87 | 12.43 | 0.12 | 0.20 | 0.22 | 0.47 |
| Ours | avg±std | 28.71±3.73 | 27.95±2.28 | 0.17±0.05 | 0.21±0.05 | 0.86±0.07 | 0.88±0.06 |
| | max | 36.52 | 33.63 | 0.34 | 0.37 | 0.96 | 0.95 |
| | min | 20.77 | 22.69 | 0.08 | 0.11 | 0.68 | 0.65 |

**Supplementary Table 4. Quantitative estimation of U-Net and our method before scaling**

| Method | | psnr_mito | psnr_micro | nrmse_mito | nrmse_micro | ssim_mito | ssim_micro |
|---|---|---|---|---|---|---|---|
| U-Net | average | 21.82509757 | 18.06861454 | 0.410831954 | 0.670568455 | 0.661112132 | 0.715127205 |
| | maximum | 32.62920616 | 27.59102334 | 0.725499649 | 1.20194712 | 0.89115484 | 0.853380494 |
| | minimum | 13.38772688 | 12.42622894 | 0.120803005 | 0.195710158 | 0.295224305 | 0.523372789 |
| | sigma | 5.771067723 | 3.95796149 | 0.183833777 | 0.268039679 | 0.158622556 | 0.065459295 |
| Ours | average | 28.02737002 | 28.25352083 | 0.184658584 | 0.196710197 | 0.829053001 | 0.865581072 |
| | maximum | 36.51861114 | 35.36722683 | 0.351261831 | 0.435220392 | 0.957965596 | 0.946455588 |
| | minimum | 20.65213506 | 22.29958129 | 0.083796356 | 0.094390796 | 0.593602829 | 0.645551091 |
| | sigma | 3.498943513 | 2.641980436 | 0.054013478 | 0.061986473 | 0.091739203 | 0.060531109 |

**Supplementary Table 5. Scaling effects on model performance**

| | Method | psnr_mito | psnr_micro | nrmse_mito | nrmse_micro | ssim_mito | ssim_micro |
|---|---|---|---|---|---|---|---|
| U-Net-diff | average | 21.83±5.77 | 18.07±3.96 | 0.41±0.18 | 0.67±0.27 | 0.66±0.16 | 0.72±0.07 |
| | maximum | 32.62920616 | 27.59102334 | 0.725499649 | 1.20194712 | 0.89115484 | 0.853380494 |
| | minimum | 13.38772688 | 12.42622894 | 0.120803005 | 0.195710158 | 0.295224305 | 0.523372789 |
| U-Net-same | average | 21.38±5.86 | 17.45±3.87 | 0.44±0.2 | 0.76±0.3 | 0.61±0.2 | 0.68±0.08 |
| | maximum | 32.62922236 | 27.5909474 | 0.766075891 | 1.36853163 | 0.891148704 | 0.840804737 |
| | minimum | 12.87002674 | 12.42597578 | 0.12080278 | 0.195711715 | 0.215861446 | 0.46885837 |
| Ours-diff | average | 28.03±3.5 | 28.25±2.64 | 0.18±0.05 | 0.20±0.06 | 0.83±0.09 | 0.87±0.06 |
| | maximum | 36.51861114 | 35.36722683 | 0.351261831 | 0.435220392 | 0.957965596 | 0.946455588 |
| | minimum | 20.65213506 | 22.29958129 | 0.083796356 | 0.094390796 | 0.593602829 | 0.645551091 |
| Ours-same | average | 28.71±3.73 | 27.95±2.28 | 0.17±0.05 | 0.21±0.05 | 0.86±0.07 | 0.88±0.06 |
| | maximum | 36.51857337 | 33.62887733 | 0.340188947 | 0.365645318 | 0.957965385 | 0.95438721 |
| | minimum | 20.77183172 | 22.68661339 | 0.083796721 | 0.106531712 | 0.680121854 | 0.645551549 |

# Supplementary figure 1 AEMS-Net architecture details

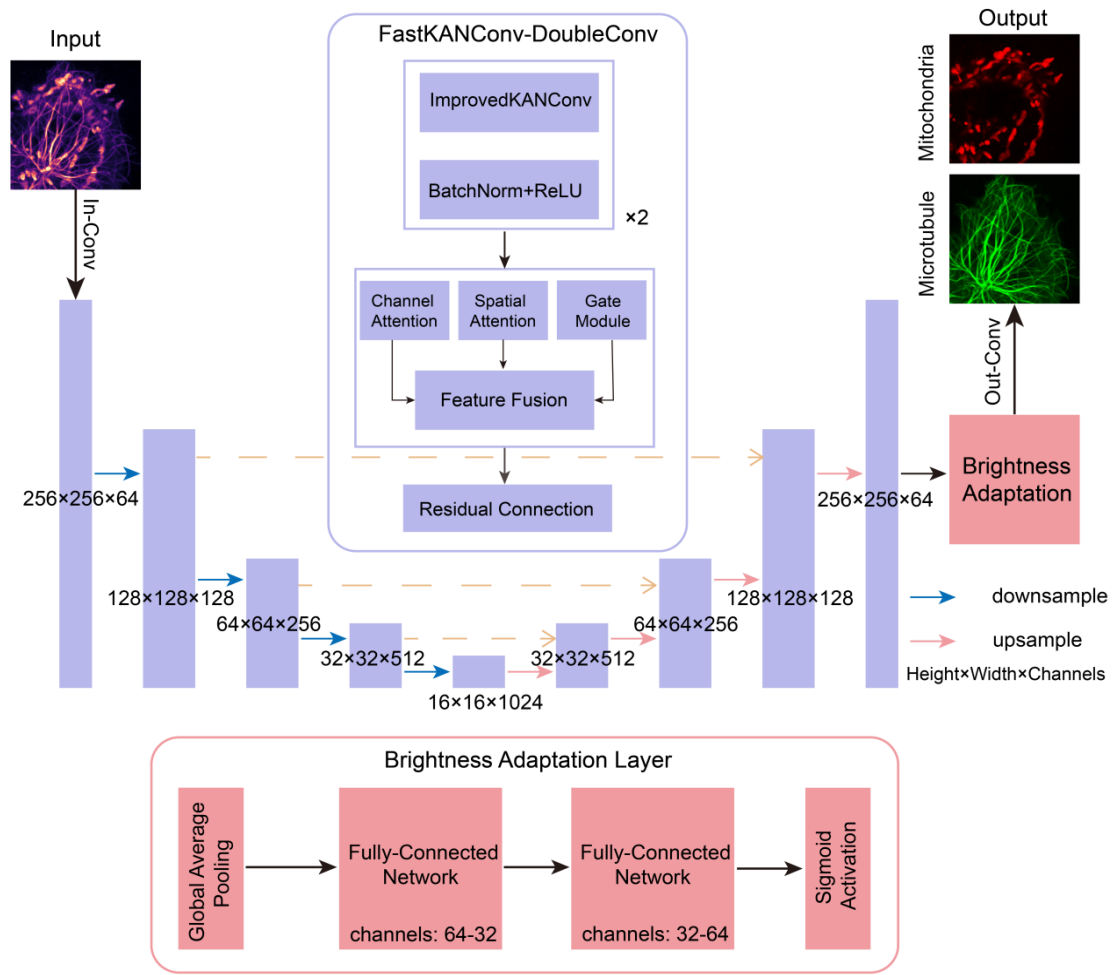

**Supplementary figure 1 AEMS-Net architecture details.** The AEMS-Net architecture integrates cascaded encoder-decoder modules with multi-dimensional attention mechanisms, encompassing channel-wise attention, spatial attention, and gating operations, followed by hierarchical feature fusion. The network incorporates a brightness adaptation layer subsequent to the terminal up sampling operation to address heterogeneous staining intensity distributions. Detailed interpretability analyses of the representative test images are presented in Supplementary Figures 5-8.

**Supplementary figure 2 KAN convolution detail.**

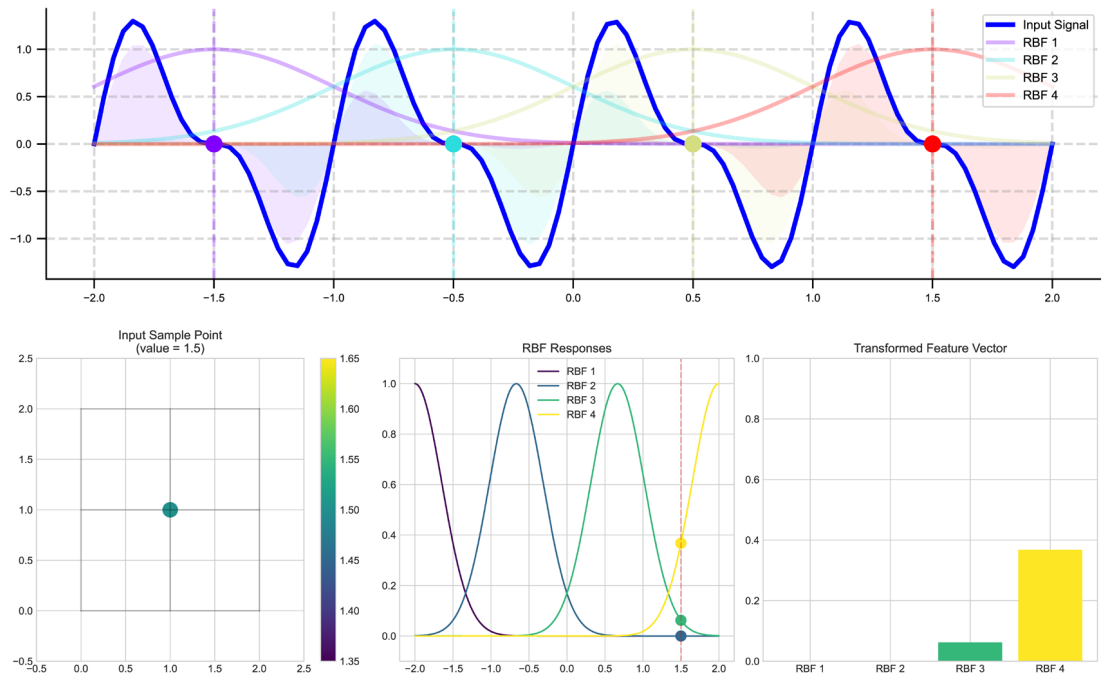

**Supplementary figure 3 KAN convolution detail.** The input dataset undergoes transformation through Radial Basis Functions, generating feature representations in elevated dimensional spaces. Subsequently, grid-based modeling followed by spline convolution operations enables precise fitting, thereby enhancing the nonlinear mapping capabilities of the computational framework.

**Supplementary figure 3 Interpretable module.**

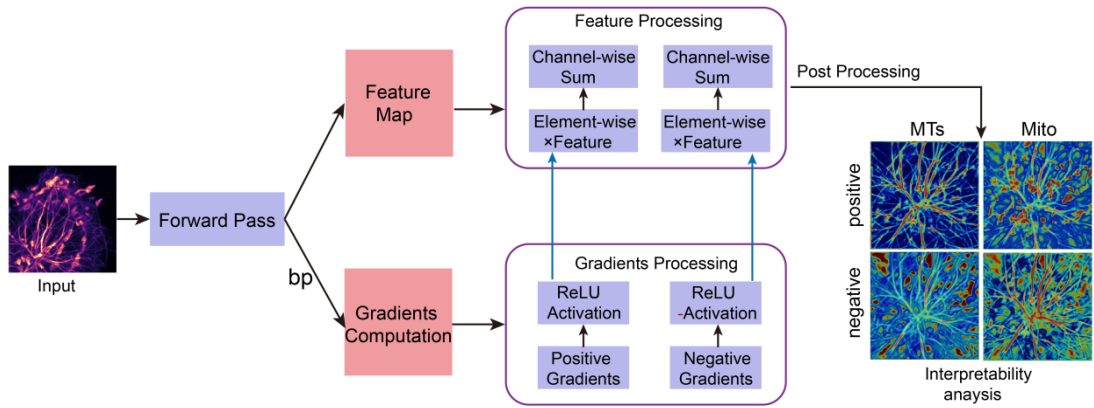

**Supplementary figure 2 Interpretable module.** Backpropagation serves as a fundamental mechanism for gradient computation within neural networks. The visualization of internal gradient propagation reveals the intricate operational dynamics of deep learning architectures. Through gradient negation, regions of minimal significance can be systematically identified and analyzed. This methodological approach facilitates comprehensive understanding of the neural decision-making mechanisms, thereby transforming the traditionally opaque nature of deep learning systems into an interpretable analytical framework.

**Supplementary Figure 4. Performance evaluation of AEMS-Net on scale-enhanced test datasets.**

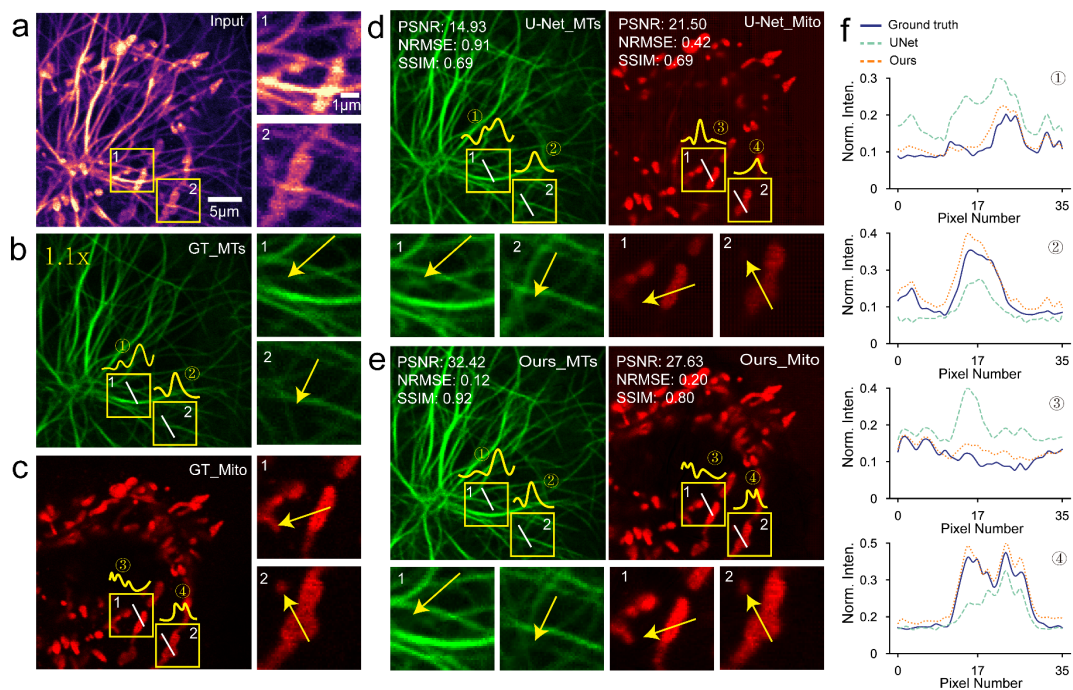

**Supplementary Figure 4. Performance evaluation of AEMS-Net on scale-enhanced test datasets. a** Overlaid test images of mitochondria and microtubules, where microtubules underwent 1.1-fold magnification at original scale. Two regions of interest (yellow boxes) were selected and magnified, with yellow arrows highlighting key features. **b-c** Ground truth images of microtubules and mitochondria. Line profiles within the selected regions generated original microtubule intensity waveforms, labeled as lines 1-4. **d-e** Reconstruction outcomes from U-Net and AEMS-Net. U-Net exhibited systematic classification errors when separating microtubules from mitochondria (panel d, yellow arrows). In contrast, AEMS-Net generated reconstructions that closely aligned with ground truth (panel e, yellow arrows). **f** Comparative analysis of normalized intensity profiles across four regions. U-Net results (green dashed lines) showed substantial deviations from original data, with classification errors in specific ranges. AEMS-Net results (orange dashed lines) demonstrated superior alignment with original data. While U-Net performance fluctuated significantly compared to non-scale-enhanced results, AEMS-Net maintained consistent performance across conditions.

**Supplementary figure 5 Time sequence of cell interact.**

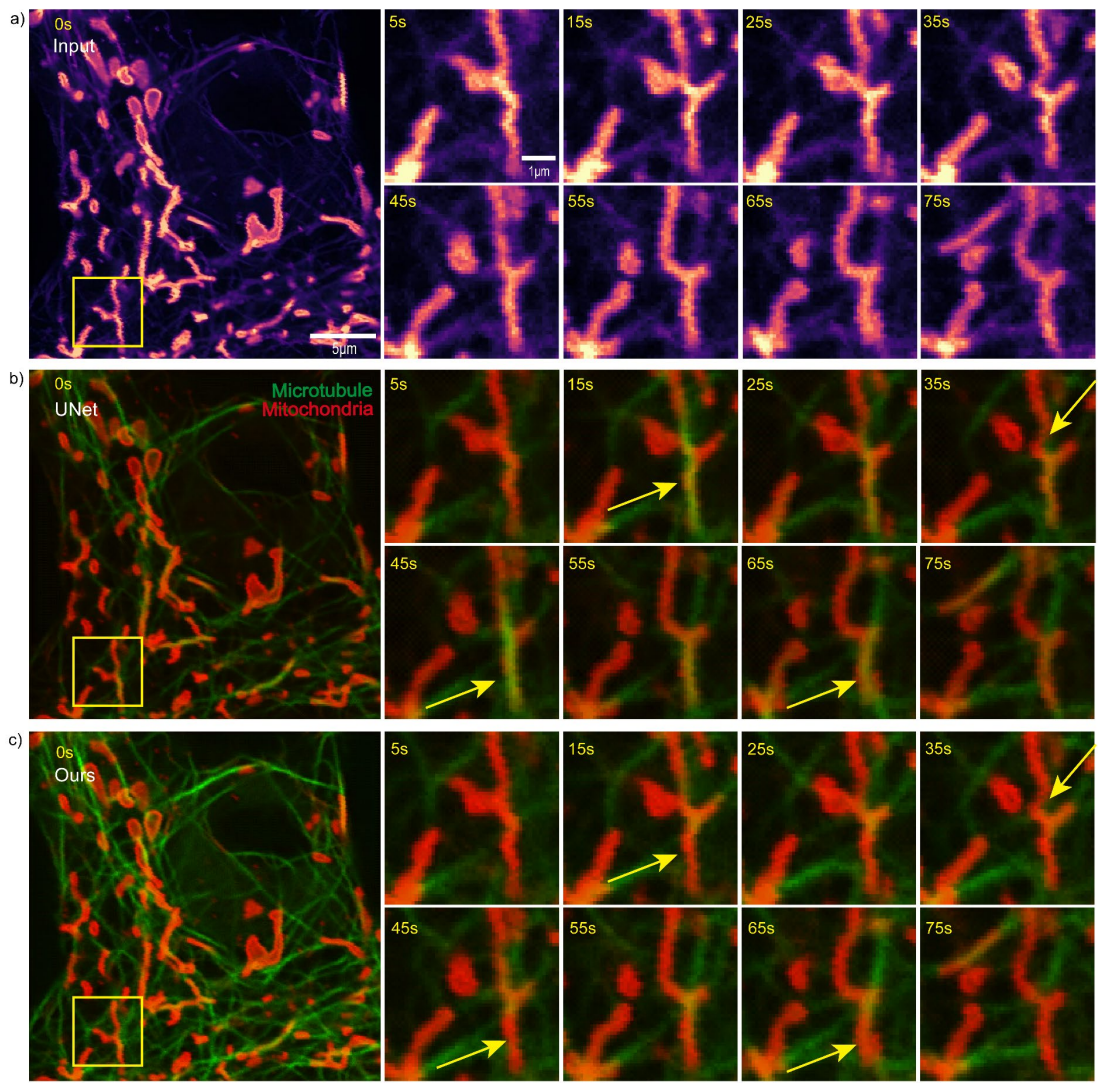

**Supplementary figure 5 Time sequence of cell interact.** Within the Application dataset, Cell1 demonstrates complex spatiotemporal dynamics. Notable divergences between the U-Net architecture and the proposed methodology manifest at multiple temporal points (15, 35, 45, and 65 seconds). The presented approach enables precise reconstruction of mitochondrial trafficking along microtubule networks, whereas the U-Net framework exhibits substantial limitations in capturing these intricate biological processes.

**Supplementary figure 6 Explanation of our method(mitochondria).**

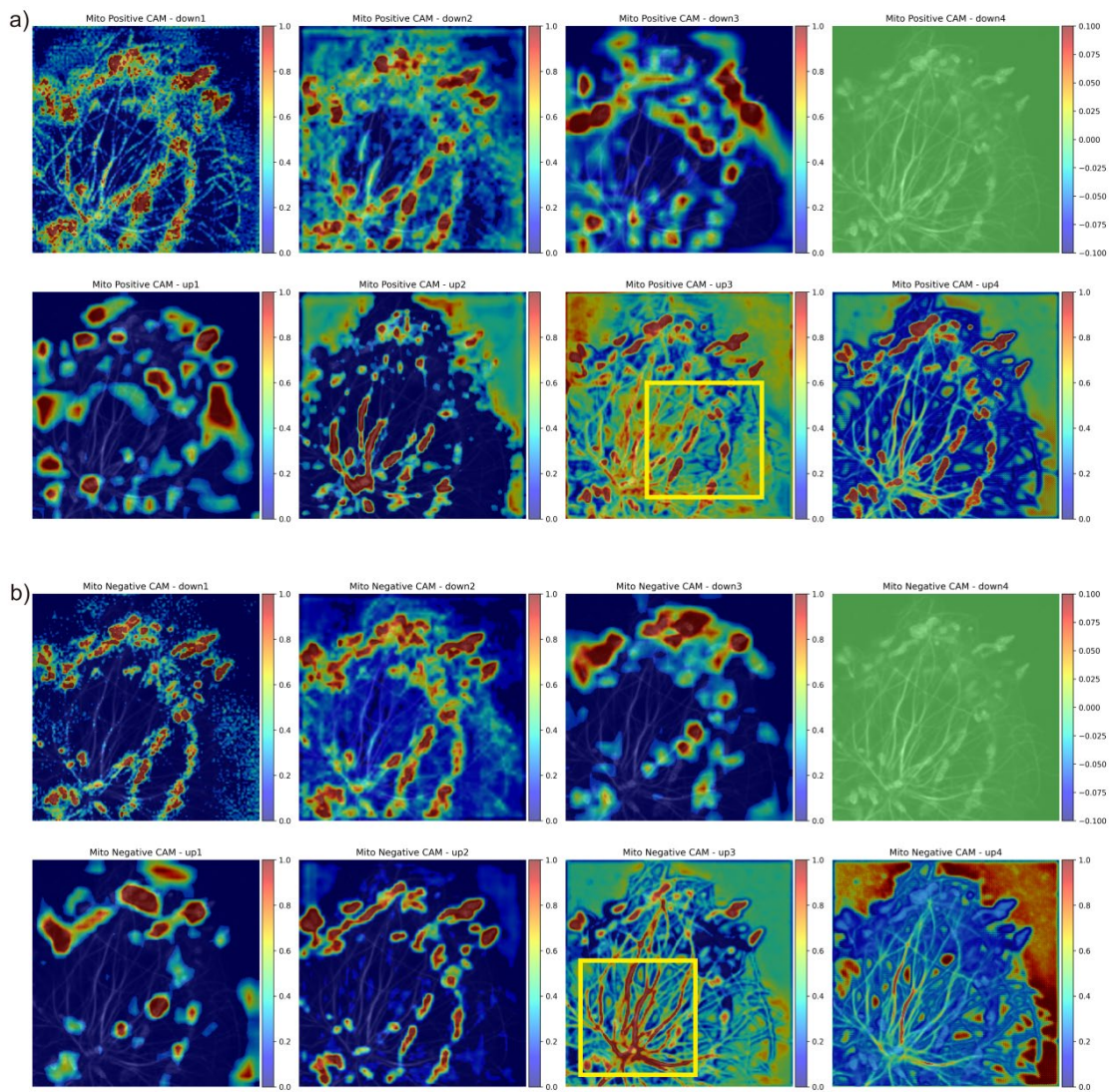

**Supplementary figure 6 Explanation of our method (mitochondria).** The interpretable analysis of mitochondrial structures through this methodology reveals complementary positive and negative activation patterns within the delineated regions of interest (highlighted in yellow). These bidirectional activation signatures demonstrate the robust capability of the analytical framework to capture antagonistic features, thereby advancing the comprehensive understanding of mitochondrial morphological characteristics and their underlying representational mechanisms within the deep learning architecture.

**Supplementary figure 7 Explanation of our method(microtubules).**

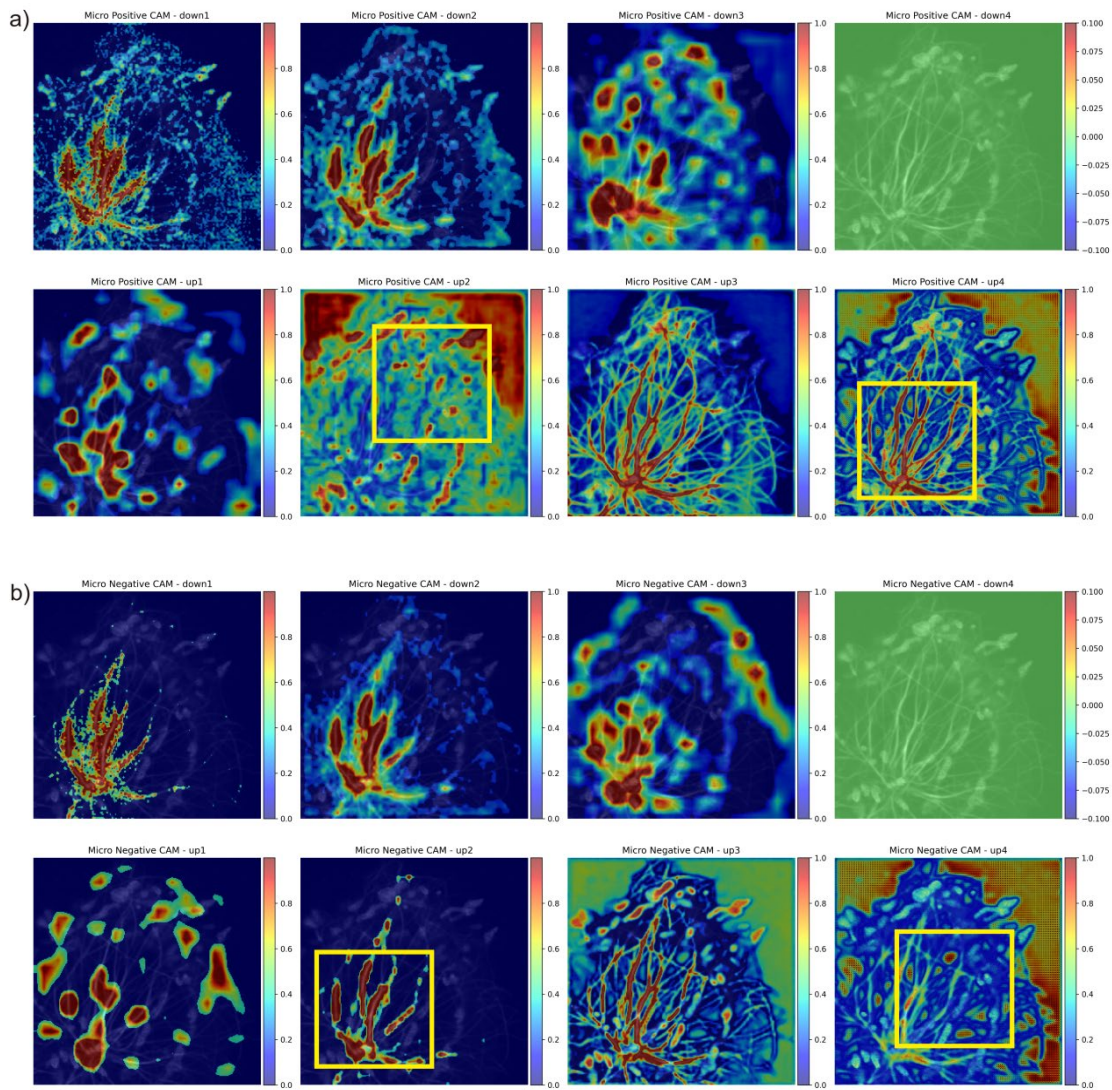

**Supplementary figure 7 Explanation of our method (microtubules).** The interpretable analysis of microtubule distributions through this methodological approach reveals distinct positive activation patterns within the demarcated yellow bounding boxes containing microtubular structures, whereas negative activation values manifest in the adjacent non-microtubular regions. This spatial contrast in activation patterns demonstrates the robust discriminative capability of the model in distinguishing microtubular architectures from other subcellular constituents.

**Supplementary figure 8 Explanation of U-Net (mitochondria).**

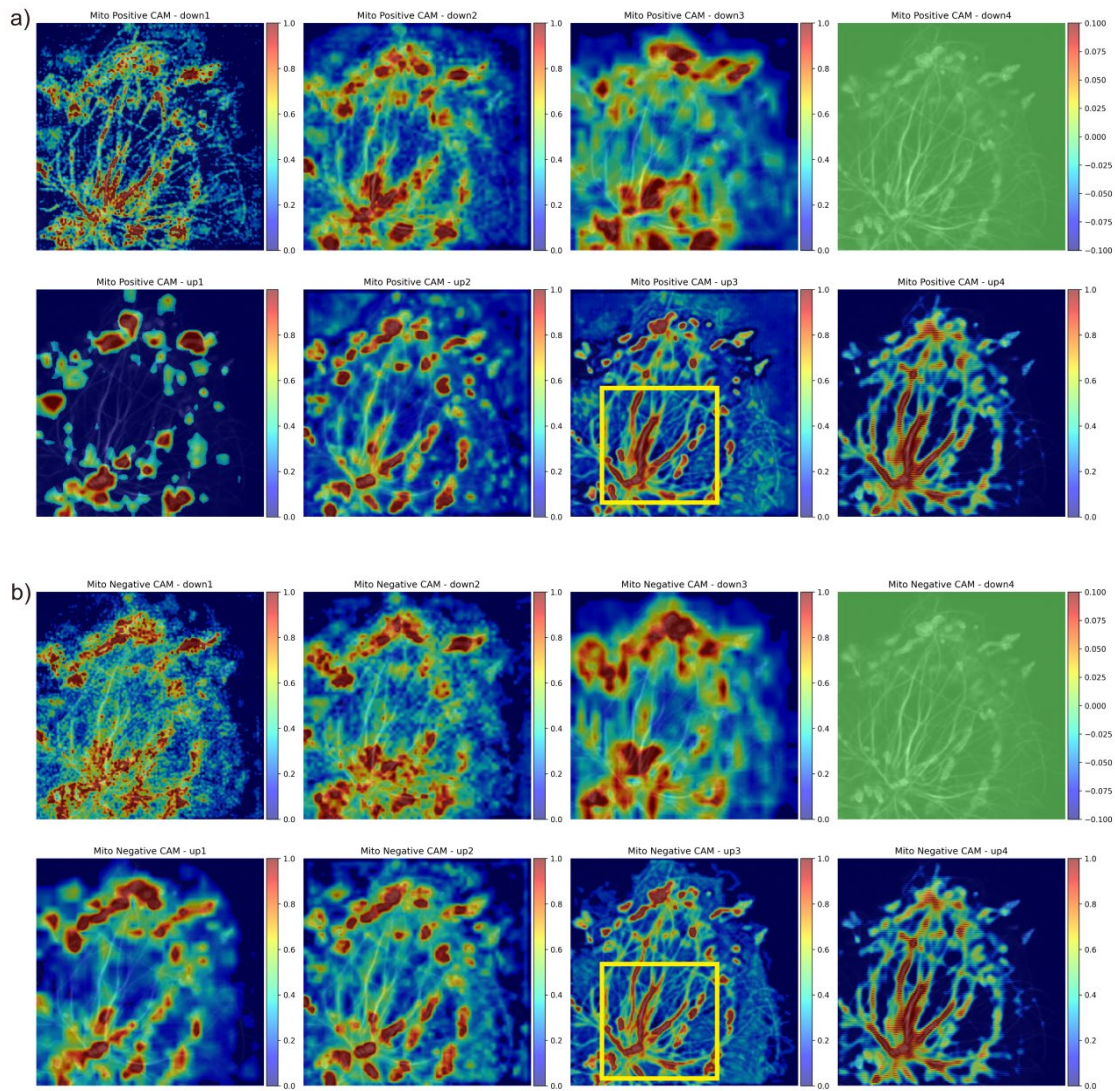

**Supplementary figure 8 Explanation of U-Net (mitochondria).** The interpretable deep learning analysis of mitochondrial ultrastructure through U-Net architecture revealed minimal differentiation in activation patterns within the designated regions of interest (demarcated by yellow bounding boxes). This observed homogeneity in activation values suggests inherent limitations in the neural network architecture regarding the effective representation and discrimination of complex mitochondrial morphological features. The findings indicate potential constraints in the computational framework for precise characterization of subcellular organelle structures.

**Supplementary figure 9 Explanation of U-Net(microtubules).**

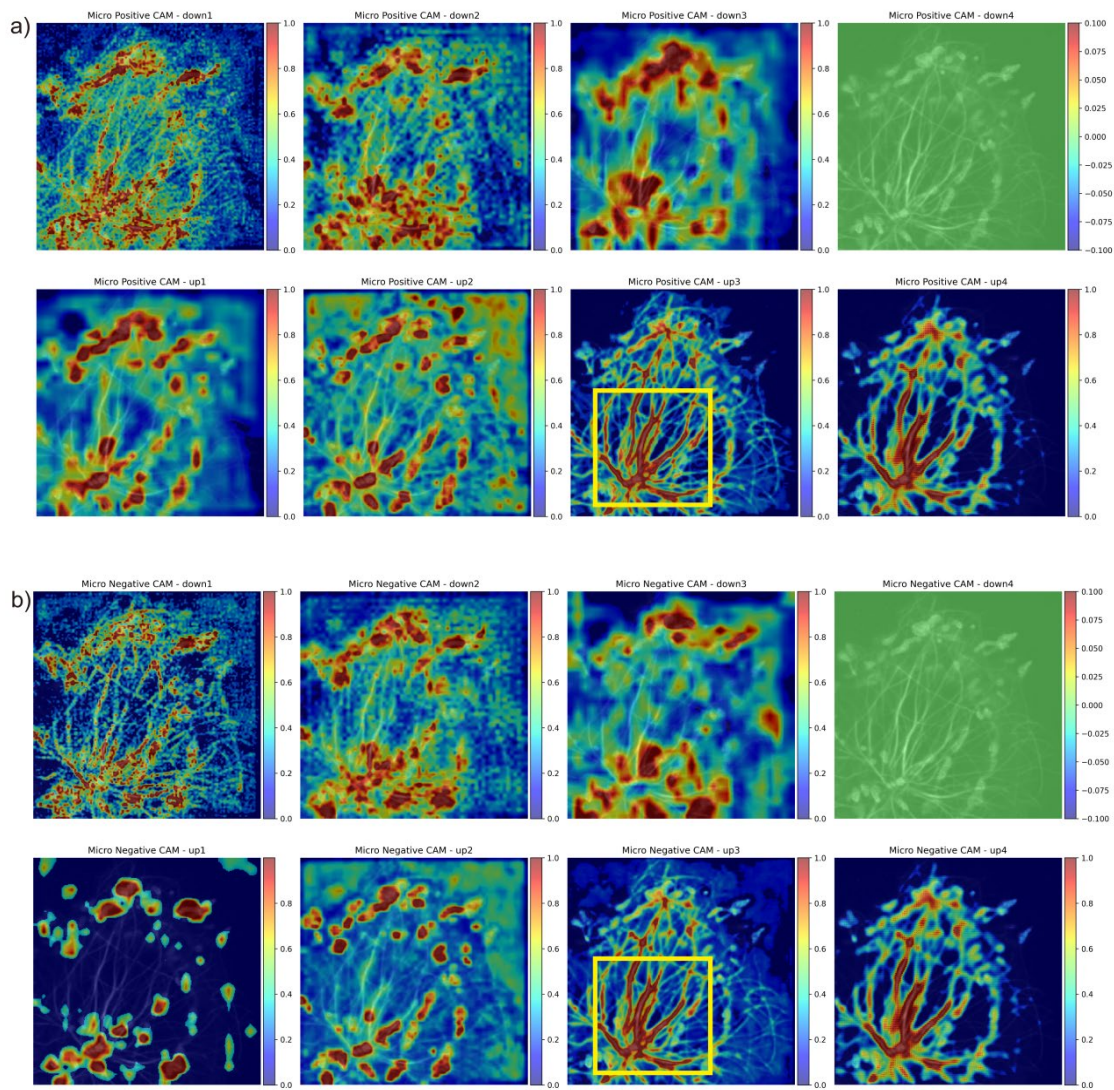

**Supplementary figure 9 Explanation of U-Net (microtubules).** The interpretable deep learning analysis of microtubular structures through U-Net architecture revealed negligible variations in activation patterns within the demarcated regions of interest. The observed homogeneity in activation values suggests limited discriminative capability of the model in characterizing and representing distinctive microtubule conformational features, indicating potential limitations in the neural network architecture for cytoskeletal pattern recognition.

**Supplementary figure 10 Comparison of our method and U-Net.**

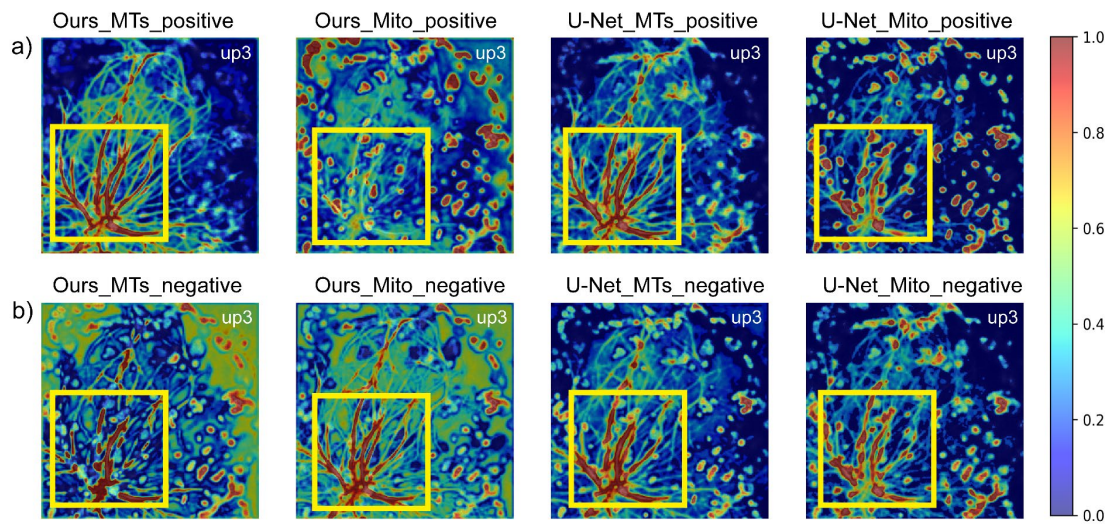

**Supplementary figure 10 Comparison of our method and U-Net.** Comparative interpretability analysis of the up3 module between our method and U-Net reveals critical differences. Our method expresses positive and negative features in alignment with human intuition, whereas U-Net demonstrates significant deviations in feature representation.

**Supplementary figure 11 Application comparison of our method and U-Net.**

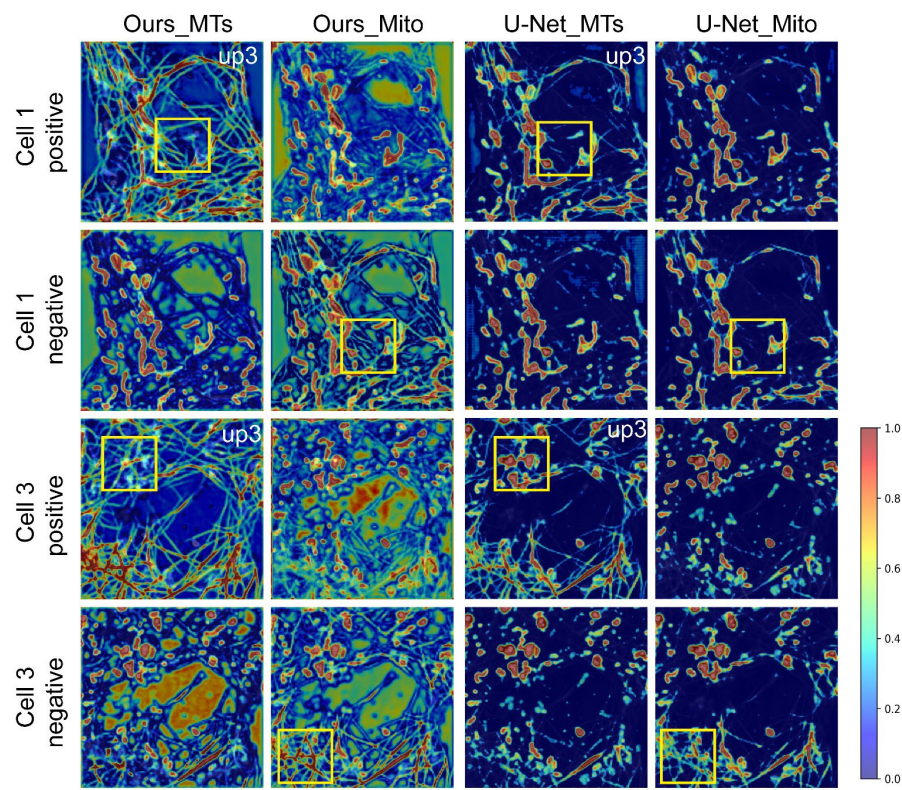

**Supplementary figure 11 Application comparison of our method and U-Net.** The proposed methodology delivers enhanced interpretability analysis in Application cells 1 and 3. The U-Net architecture exhibits limitations in discriminating between microtubular structures and mitochondrial components effectively. The regions demarcated by yellow boundaries demonstrate the enhanced analytical capabilities and superior interpretability afforded by this novel approach.